\documentclass[sigconf]{acmart}

\renewcommand\footnotetextcopyrightpermission[1]{} % removes footnote with conference information in first column
\usepackage{xspace}
\usepackage{xcolor,color,soul}
\usepackage{tabularx}
\usepackage{tcolorbox}
\usepackage{subcaption}
\tcbuselibrary{listings,breakable}

\AtBeginDocument{%
  }
\setcopyright{acmlicensed}
\copyrightyear{2018}
\acmYear{2018}
\acmDOI{XXXXXXX.XXXXXXX}
\acmConference[Conference acronym 'XX]{Make sure to enter the correct
  conference title from your rights confirmation email}{June 03--05,
  2018}{Woodstock, NY}
\acmISBN{978-1-4503-XXXX-X/2018/06}

\begin{document}

\newcommand{\paperexec}{PaperRepro\xspace}

\title{\paperexec: Automated Computational Reproducibility Assessment for Social Science Papers}

\author{Linhao Zhang}
\affiliation{
  \institution{Shandong University, Zhongguancun Academy, China}
  \state{}
  \country{}
}
% \affiliation{
%   \institution{Zhongguancun Academy}
%   \state{Beijing}
%   \country{China}
% }

\author{Tong Xia*}
\affiliation{
  \institution{Tsinghua University}
  \state{Beijing}
  \country{China}
}

\author{Jinghua Piao}
\affiliation{
  \institution{Tsinghua University}
  \state{Beijing}
  \country{China}
}

\author{Lizhen Cui}
\affiliation{
  \institution{Shandong University}
  \state{Jinan}
  \country{China}
}

\author{Yong Li*}
\affiliation{
  \institution{Tsinghua University, Zhongguancun Academy, China}
  \state{}
  \country{}
  \authornote{Corresponding authors: \{tongxia, liyong07\}@tsinghua.edu.cn}
}
% \affiliation{
%   \institution{}
%   \state{}
%   \country{China}
% \newcommand{\paperexec}{\text{Paper2Exec}}
% \newcommand{\paperexec}{Paper2Exec\xspace}
% \newcommand{\codev}{\textsc{CodeV}\xspace}
\newcommand{\tx}[1]{\sethlcolor{pink}\hl{[Tong: #1]}}

\begin{abstract}
% Computational reproducibility is a fundamental requirement for credible social science, yet scalable assessment remains challenging due to the execution-centric nature of reproducing and verifying result artifacts. Recent advances in large models have inspired agent-based approaches, but existing systems often break down due to single-context bottlenecks, tooling inadequacy, and GUI output invisibility. \tx{To address these crucial problems},  We propose a  \tx{novel} two-stage, file-driven multi-agent framework that separates artifact-generating execution from evidence-grounded evaluation, enabling traceable workflows and reliable comparison between reproduced outputs and reported results. On an established social science reproducibility assessment benchmark, our approach achieves state-of-the-art accuracy \tx{explain what is accuracy} with a 21.9\% relative gain over the strongest baseline. \tx{furthermore} We also introduce a refined and difficulty-stratified benchmark, REPRO-Bench-S, to support more diagnostic evaluation and accelerate progress in automated reproducibility assessment.

Computational reproducibility is essential for the credibility of scientific findings, particularly in the social sciences, where findings often inform real-world decisions. Manual reproducibility assessment is costly and time-consuming, as it is nontrivial to reproduce the reported findings using the authors' released code and data. Recent advances in large models (LMs) have inspired agent-based approaches for automated reproducibility assessment. However, existing approaches often struggle due to limited context capacity, inadequate task-specific tooling, and insufficient result capture. To address these, we propose \paperexec, a novel two-stage, multi-agent approach that separates execution from evaluation. In the execution stage, agents execute the reproduction package and edit the code to capture reproduced results as explicit artifacts. In the evaluation stage, agents evaluate reproducibility using explicit evidence. \paperexec assigns distinct responsibilities to agents and equips them with task-specific tools and expert prompts, mitigating context and tooling limitations. It further maximizes the LM’s coding capability to enable more complete result capture for evaluation.
On REPRO-Bench, a social science reproducibility assessment benchmark, \paperexec achieves the best overall performance, with a 21.9\% relative improvement in score-agreement accuracy over the strongest prior baseline. We further refine the benchmark and introduce REPRO-Bench-S, a benchmark stratified by execution difficulty for more diagnostic evaluation of automated reproducibility assessment systems. Our code and data are publicly available\footnote{\url{https://github.com/luolin101/PaperRepro}}.
\end{abstract}





\maketitle

\section{Introduction}
% Computational reproducibility is a fundamental prerequisite for credible scientific discovery. Scientific progress relies not only on generating new findings but also on verifying that reported results can be independently reproduced from the underlying data and code. In the social sciences, reproducibility is particularly critical, as empirical findings directly influence public policy, institutional design, and societal decision-making. Recent advances in large models (LMs) have significantly accelerated the production of scientific artifacts, including empirical analyses, code, and full research papers. This shift has lowered the barrier to producing computational social science research, while placing greater demands on reproducibility assessment. Without scalable mechanisms for assessing reproducibility, errors and inconsistencies can propagate more rapidly, undermining the reliability of scientific knowledge. As a result, automated reproducibility assessment has become essential for sustaining and accelerating social science research.

Computational reproducibility is a fundamental prerequisite for credible scientific discovery~\cite{peng2011reproducible}. Scientific progress relies not only on generating new findings but also on verifying that reported results can be independently reproduced from the underlying data and code~\cite{national2019reproducibility}. In the social sciences, reproducibility is particularly critical, as empirical findings directly influence public policy, institutional design, and societal decision-making~\cite{miguel2014promoting,hardwicke2020empirical}. Recent advances in AI have significantly accelerated the production of research papers~\cite{kusumegi2025scientific,liang2025quantifying,hao2026artificial}, while simultaneously increasing the burden of reproducibility assessment as a growing number of social science findings may be difficult to validate. 
There is a growing concern that, without scalable approaches for assessing reproducibility, errors and inconsistencies can propagate more rapidly, undermining the reliability of scientific knowledge~\cite{richardson2025entities,freedman2015economics}. 
% As a result, automated reproducibility assessment has become essential for sustaining and accelerating \tx{strengthening} social science research.
As a result, automated reproducibility assessment has become essential for strengthening social science research.
% These trends motivate the need for automated approaches for reproducibility assessment in the social sciences.

% Automating reproducibility assessment in social science is substantially more challenging than in domains such as machine learning. First, social science workflows predominantly rely on R and Stata rather than Python, increasing the difficulty of code understanding, generation, and execution due to weaker standardization and heavy use of domain-specific statistical libraries. Second, reproduction targets are typically tables and figures instead of scalar metrics, making logical inspection insufficient: valid assessment requires executing code and performing multimodal comparisons between reproduced outputs and reported results. Third, reproduction packages often lack consistent file organization or naming conventions, requiring agents to infer execution order and file dependencies from unstructured documentation. Finally, environment setup is rarely standardized, with scripts assuming implicit package versions, operating systems, or local paths, further complicating automated execution.

\begin{figure}[t]
  \includegraphics[width=\columnwidth]{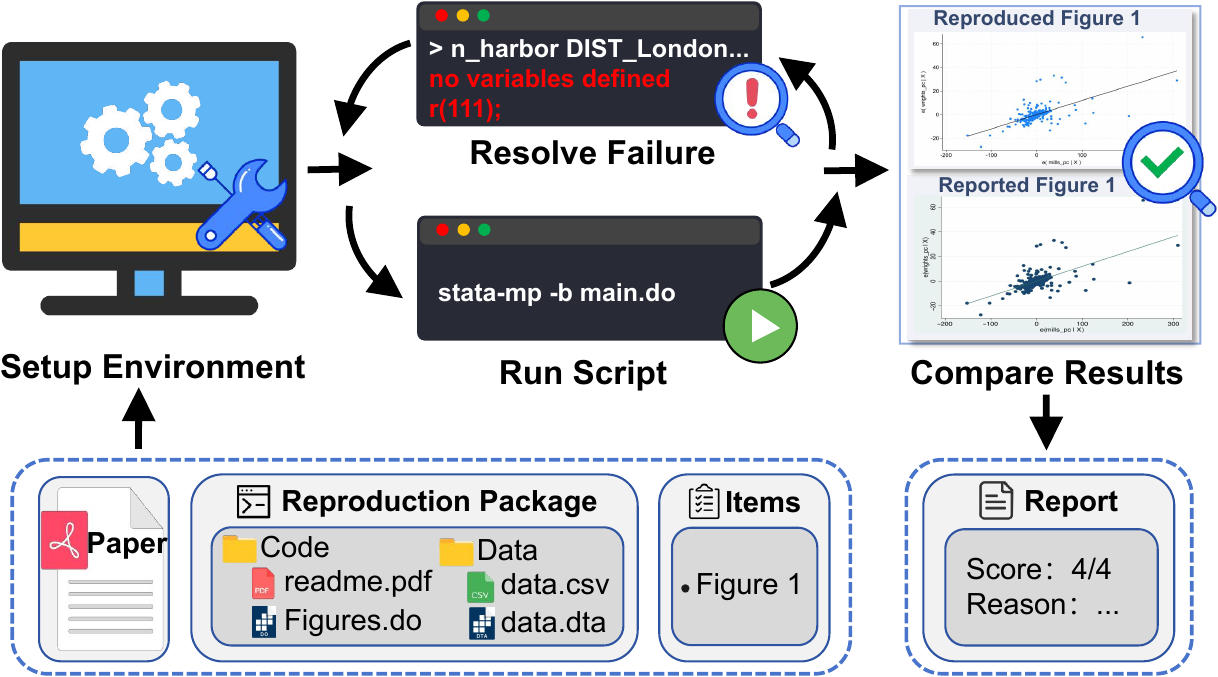}
  \caption{Pipeline of computational reproducibility assessment. A challenging process of executing the reproduction package and verifying the reproduced results against the paper’s reported results.}
  \label{fig:workflow}
\end{figure}

% Manually assessing the computational reproducibility of social science papers is costly and time-consuming. 
However, reproducibility assessment in the social sciences is still largely performed manually, which is costly and time-consuming.
For example, reproducing 110 social science papers involved 347 researchers, and a psychology replication project took over five years for 100 studies~\cite{RePEc:zbw:i4rdps:107, project_wiki}. This burden largely stems from the complexity of the assessment process.
% As shown in Figure~\ref{fig:workflow}, given the paper, the reproduction package, and reproduction items (e.g., tables or figures), reproducibility assessment aims to produce a reproducibility score and typically involves setting up the environment, running scripts, resolving failures, and comparing reproduced results with the paper’s reported results. 
As shown in Figure~\ref{fig:workflow}, given a paper, a reproduction package with released code and data, and reproduction items (e.g., tables or figures), reproducibility assessment aims to produce a reproducibility score. It typically involves setting up the environment, running scripts, resolving failures, and comparing reproduced results with the paper’s reported results.
In practice, run instructions are often unclear, dependencies and software versions are often implicit, workflows may span multiple tools (e.g., R, Stata) and domain-specific libraries, and verification frequently involves tables and figures. These practical challenges make this process difficult to automate, which we analyze in more detail in Section~\ref{sec:background}.

Recent advances in large models (LMs) have demonstrated strong capabilities in code understanding, generation, and tool use~\cite{jiang2024survey,zhang-etal-2024-codeagent}, offering new hope for automating reproducibility assessment. Accordingly, recent work has proposed agent-based approaches to assess reproducibility~\cite{siegel2024core,hu2025repro}. 
Despite this progress, existing approaches still struggle in the social sciences due to three gaps.
First, they face a \textbf{context bottleneck}. In social science reproduction, agents must track many files, scripts, logs, and intermediate outputs. This volume of information can exceed the agent’s context capacity and cause critical details to be dropped. 
Second, they exhibit a \textbf{task-specific tooling gap}. They mostly rely on generic command execution and basic file operations, which provide limited support for the social science reproduction process centered on document and result artifacts. For example, agents cannot directly extract a specific table or figure from a paper for comparison and instead resort to page-level search over the entire PDF, which becomes especially brittle for long documents. 
Third, they suffer from a \textbf{result capture gap}. Existing agents often focus on running the original scripts while overlooking the capture of reproduced results. Results may appear only in graphical or interactive interfaces, in runtime logs, or as multiple tables and figures, which makes it hard to identify the intended result for verification.

To address these challenges, we propose \paperexec, a two-stage, multi-agent approach for automated reproducibility assessment of social science papers. It consists of an execution stage and an evaluation stage. 
In the artifact-driven execution stage, a Setup Agent configures the environment and produces a reproduction plan. An Execution Agent follows the plan to run scripts, editing the code to capture reproduced results and persist them as reproduced artifacts.
In the evidence-grounded evaluation stage, a Scoring Agent evaluates reproducibility using explicit evidence, including reproduced artifacts, the paper’s reported results, and an execution summary. A Report Agent aggregates the execution trace and evaluation outcomes into a structured report.
We split the assessment process into two stages and further decompose it into subtasks assigned to different agents. Stage separation reduces interference from execution traces during evaluation, making the assessment more reliable. Task decomposition keeps each agent focused on a single responsibility and alleviates context pressure. We further equip each agent with task-specific tools and expert prompts, specializing agents for their assigned task. Finally, \paperexec adopts a file-driven interaction design. Agents write task deliverables to files, producing a traceable and verifiable assessment record that supports human review.

We conduct extensive experiments to evaluate \paperexec. On REPRO-Bench, a benchmark for social science reproducibility assessment, \paperexec achieves the best overall performance. In particular, it improves accuracy by 21.9\% relative to the best prior baseline, where accuracy measures score agreement. We further perform ablations to analyze the impact of task decomposition and specialized tools, and examine failure cases to identify remaining challenges and limitations. To assess generalization beyond the benchmark, we present a case study on a \textit{Nature} paper that illustrates \paperexec’s end-to-end assessment process. 
% Beyond evaluating \paperexec, we refine REPRO-Bench by correcting identified issues and introducing REPRO-Bench-S, a difficulty-stratified version designed to support future research on reproducibility assessment in the social sciences.
During failure analysis, we observe a small number of annotation issues in REPRO-Bench. We therefore refine the benchmark by correcting the identified issues, and further introduce REPRO-Bench-S, a difficulty-stratified version designed to support future research on reproducibility assessment in the social sciences.
Overall, the contributions of this paper are as follows:
\begin{itemize}
\item We propose a multi-agent approach for automatically assessing the computational reproducibility of social science papers. It equips agents with task-specific tools and expert prompts to specialize their capabilities, and uses file-based deliverables for traceability and human review.

% \item We design a two-stage pipeline that improves assessment reliability by separating execution from evaluation. In the artifact-driven execution stage, agents execute the repository and capture reproduced results as artifacts. In the evidence-grounded evaluation stage, agents evaluate reproducibility using explicit evidence and generate a report.
\item We design a two-stage pipeline that reduces interference from lengthy execution traces during evaluation, improving assessment reliability. 
% The execution stage executes the repository and captures reproduced results, while the evaluation stage evaluates reproducibility based on evidence. \tx{The second sentence can be removed}

\item We conduct extensive experiments to evaluate \paperexec. On REPRO-Bench, \paperexec achieves the best performance and improves accuracy by 21.9\% relative to the best prior baseline.

\item We refine REPRO-Bench by correcting issues and introducing REPRO-Bench-S, a difficulty-stratified benchmark that enables more precise and diagnostic evaluation of automated reproducibility assessment systems.
\end{itemize}

\section{Background}
\label{sec:background}
\paragraph{Problem definition.} In this work, we study the problem of computational reproducibility assessment in social science research. Given a paper, its reproduction package with released code and data, and a set of reproduction items, the task is to determine whether the reported results can be reproduced using the same computational procedures, and assign a reproducibility score.

\begin{figure}[t]
  \includegraphics[width=\columnwidth]{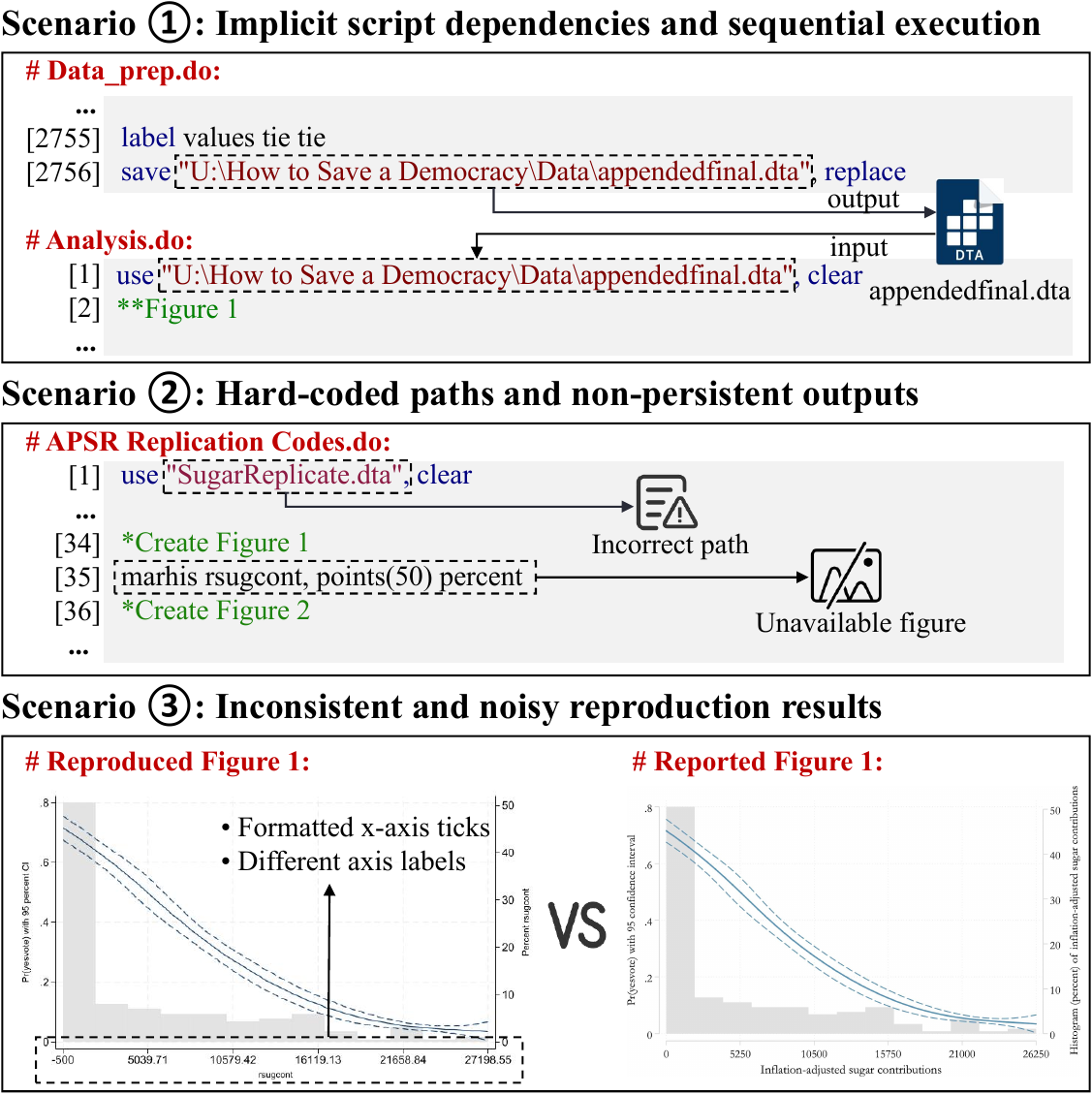}
  \caption{Three representative reproduction scenarios that commonly hinder automated reproducibility assessment in social science.}
  \label{fig:challenges}
\end{figure}

\paragraph{Practical challenges for automation.} Automated reproducibility assessment requires executing the reproduction package and comparing reproduced results with the paper, which remains challenging in practice. Figure~\ref{fig:challenges} highlights three common reproduction scenarios in social science research that frequently cause automation to fail. These scenarios are distilled from cases in REPRO-Bench~\cite{hu2025repro}, which is largely built on a large-scale reproduction effort covering 110 papers in leading economic and political science journals~\cite{RePEc:zbw:i4rdps:107}, making our selected scenarios representative.

\textbf{Scenario \textcircled{1}: Implicit script dependencies and sequential execution.}
Social science reproduction packages often contain multiple analysis scripts that depend on files produced by other scripts, so running scripts independently or in an arbitrary order can fail. For instance, one script first generates an intermediate dataset (e.g., \texttt{appendedfinal.dta}), and a later analysis script loads this file to produce the target figure. If the latter script is executed too early, it may fail because the intermediate file is not yet available, or it may produce incorrect outputs. The difficulty is that the intended order is not always stated explicitly and must be inferred from file input/output statements and the repository structure.

\textbf{Scenario \textcircled{2}: Hard-coded paths and non-persistent outputs.} Reproduction scripts often hard-code local file paths, which causes execution to fail when the package is run in a new environment. For instance, a script may load a dataset from an absolute path on the authors’ machine. The system must rewrite the path to point to the correct local location. Otherwise, the script fails and produces no results. 
Even when execution succeeds, the target outputs may still be unavailable to automated assessment. Some scripts only display figures on screen during execution and do not save them as files. In this case, the system cannot access the reproduced artifact for later comparison. 

\textbf{Scenario \textcircled{3}: Inconsistent and noisy reproduction results.} Even when execution succeeds, reproduced artifacts may not exactly match those reported in the paper. For instance, a reproduced figure may have different axis labels, tick formatting, or visual presentation than the reported figure. Similarly, reproduced tables may include extra rows, different naming conventions, or small numerical differences due to rounding. These discrepancies require careful alignment between reproduced artifacts and reported results, rather than direct equality checks.

Beyond these three scenarios, many other challenges arise in practice, such as inferring which scripts reproduce a given item and identifying the corresponding artifacts among many outputs. Existing agent-based approaches for reproducibility assessment~\cite{siegel2024core,hu2025repro} still struggle with these challenges.

\section{\paperexec}
\begin{figure*}[t]
  \includegraphics[width=2\columnwidth]{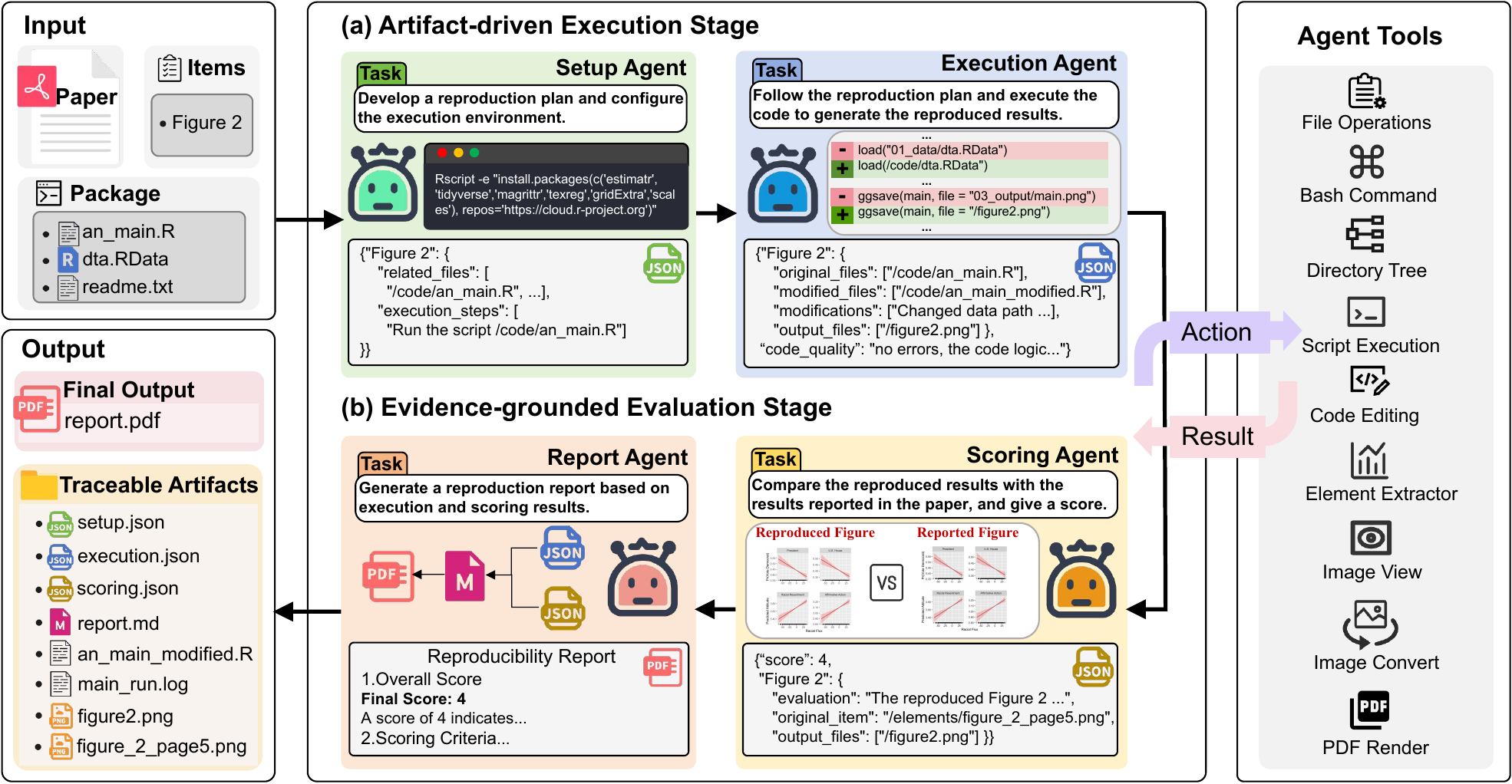}
  \caption{Overview of \paperexec. It follows a two-stage pipeline: (a) an artifact-driven execution stage that executes the reproduction package and captures reproduced artifacts, and (b) an evidence-grounded evaluation stage that aligns artifacts with the paper to produce the final score and report.}

  \label{fig:framework}
\end{figure*}

% In this section, we present \paperexec. As shown in Figure~\ref{fig:framework}, \paperexec is a multi-agent framework with a two-stage pipeline. In the artifact-driven execution stage, agents execute the reproduction package and save the reproduced artifacts for later inspection. In the evidence-grounded evaluation stage, agents evaluate reproducibility based on the reproduced artifacts and the target paper, producing a final score and report. To reliably perform execution and evaluation, \paperexec equips agents with a set of tools. We next describe the two stages and the key designs of \paperexec.

% In this section, we present \paperexec. As shown in Figure~\ref{fig:framework}, \paperexec is a multi-agent framework with a two-stage pipeline. The \emph{artifact-driven execution stage} executes the reproduction package and persistently records reproduction artifacts. The \emph{evidence-grounded evaluation stage} aligns reproduced artifacts with the paper and produces the final score and report. To support reliable interaction with the repository and execution environment, \paperexec equips agents with a set of tools. We next describe the two stages and the key designs of \paperexec.

% Reproducibility assessment aims to generate an overall score $S$ given a paper $P$, its reproduction package $R$, and reproduction items $\mathcal{I}$. 
In this section, we present \paperexec. As shown in Figure~\ref{fig:framework}, \paperexec is a multi-agent approach with a two-stage pipeline. Given a paper $P$, its reproduction package $C$, and reproduction items $\mathcal{I}$, the \emph{artifact-driven execution stage} executes $C$, captures the reproduced artifacts $\mathcal{A}_{r}$, and produces an execution summary $\mathcal{D}_{e}$.
The \emph{evidence-grounded evaluation stage} aligns $\mathcal{A}_{r}$ with $P$, then references $\mathcal{D}_{e}$ to produce the reproducibility score $S$ and report $R$.
Formally, the overall process is defined as:
\begin{equation}
\mathcal{A}_{r}, \mathcal{D}_e=\mathcal{M}_{\mathsf{Exec}}(P,C,\mathcal{I}), \quad S,R=\mathcal{M}_{\mathsf{Eval}}(P,C,\mathcal{I},\mathcal{A}_{r},\mathcal{D}_{e})
\end{equation}
where $\mathcal{M}_{\mathsf{Exec}}$ and $\mathcal{M}_{\mathsf{Eval}}$ denote the two stages. \paperexec equips agents with a set of tools to interact with the environment, improving their ability to complete the reproducibility assessment process. We next describe each stage and the key designs of \paperexec.

\subsection{Artifact-driven Execution Stage}
This stage executes the reproduction package $C$ and produces reproduced artifacts $\mathcal{A}_{r}$ for the reproduction items $\mathcal{I}$. To handle this complex process, \paperexec splits it into two subtasks assigned to two agents. The Setup Agent focuses on planning and preparation, while the Execution Agent focuses on execution and artifact capture.

\textbf{Setup Agent.} This agent prepares the reproduction environment and constructs an executable reproduction plan $\mathcal{D}_{p}$. We denote its procedure by $\mathsf{A_{set}}$:
\begin{equation}
\mathcal{D}_{p}=\mathsf{A_{set}}(P,C,\mathcal{I})
\end{equation}
Specifically, it inspects the repository, including documentation (e.g., README files) and source code, to identify entry scripts relevant to each reproduction item. Based on these entry points, it infers dependencies among scripts, determines a valid execution order, and encodes the ordered commands into $\mathcal{D}_{p}$ together with required auxiliary resources, such as input datasets and related scripts. For environment setup, it identifies required dependencies from scripts and documentation, then consolidates them into a single installation script. This avoids the resource overhead of multi-round installation and reduces the risk of missing dependencies.

\textbf{Execution Agent.} This agent follows the reproduction plan $\mathcal{D}_{p}$ to run the scripts and capture the reproduced artifacts $\mathcal{A}_{r}$. We denote its procedure by $\mathsf{A_{exe}}$:
\begin{equation}
\mathcal{A}_{r}, \mathcal{D}_{e}=\mathsf{A_{exe}}(P,C,\mathcal{I},\mathcal{D}_{p})
\end{equation}
Specifically, it reads the scripts and applies minimal, semantics-guided edits to persist relevant outputs as artifacts in $\mathcal{A}_{r}$, which is particularly useful when results are only shown in GUIs or logs. When execution errors occur, it iteratively debugs failures via log inspection and minimal script edits. To preserve traceability, it applies edits to copied scripts rather than overwriting the originals. It also performs a lightweight semantic inspection to assess code quality and detect potential logical or coding errors, consulting the paper $P$ when needed. After execution completes, it produces $\mathcal{D}_{e}$, which records the modifications and generated artifacts for each reproduction item, together with an overall code quality assessment for the reproduction package.

\textbf{Agent Tools.} Both agents share three general-purpose tools for repository interaction, including (1) file operations, (2) directory inspection, and (3) bash commands. Each agent is also equipped with task-specific tools. The Setup Agent uses a Python execution tool inspired by CodeAct~\cite{wang2024executable} to install dependencies in a controlled and unified manner. The Execution Agent uses specific tools for robust script execution and iterative debugging, including (1) multi-language script runners that execute scripts and capture logs automatically, and (2) a lightweight editing tool that supports content-based search and replacement for minimal, traceable edits.

\subsection{Evidence-grounded Evaluation Stage}
This stage evaluates reproducibility based on the reproduced artifacts $\mathcal{A}_{r}$, the paper $P$, and the execution summary $\mathcal{D}_{e}$. It outputs the score $S$ and the report $R$. \paperexec assigns two subtasks to two agents. The Scoring Agent focuses on evidence-grounded scoring, while the Report Agent focuses on report generation.

\textbf{Scoring Agent.} This agent gathers evidence for each reproduction item and assigns a reproducibility score. We denote its procedure by $\mathsf{A_{scr}}$:
\begin{equation}
S,\mathcal{D}_{s}=\mathsf{A}_\mathsf{scr}(P,C,\mathcal{I},\mathcal{A}_{r},\mathcal{D}_{e})
\end{equation}
Specifically, for each item, it (i) retrieves the corresponding reported result in the paper $P$ (e.g., a figure or table), (ii) locates the reproduced artifacts in $\mathcal{A}_{r}$ using the records in the execution summary $\mathcal{D}_{e}$, and (iii) when expected artifacts are missing, reconstructs the reproduced result from execution evidence (e.g., logs and intermediate files). It then builds an evidence set that includes the reported results, the reproduced artifacts, and the code quality assessment recorded in $\mathcal{D}_{e}$. Based on this evidence, it evaluates consistency between reproduced results and reported results, then assigns $S$ by combining item-level consistency with the code quality assessment. The agent outputs a scoring summary $\mathcal{D}_{s}$, which records the reported results, reproduced artifacts, and consistency assessments for each item, together with the overall score.

\textbf{Report Agent.}
This agent produces a structured reproducibility report by consolidating the execution and scoring summaries. 
We denote its procedure by $\mathsf{A_{rep}}$: 
\begin{equation}
R=\mathsf{A_{rep}}(P,C,\mathcal{I},\mathcal{D}_{e},\mathcal{D}_{s})
\end{equation}
Specifically, it uses predefined templates to organize the overall score, scoring criteria, overall assessment, and per-item results into a coherent document. The report summarizes key reproduction steps, major successes and failure points, and the evidence supporting the final score. It outputs both machine-readable records and a human-readable PDF for inspection and archival.

\textbf{Agent Tools.} 
Both agents inherit the three general-purpose repository tools described in the execution stage. In addition, the Scoring Agent, backed by a multimodal model, uses task-specific tools for multimodal verification, including (1) an element extractor that exports tables and figures from the paper as images in their original order, (2) an image viewing tool that brings visual content into the agent’s context, and (3) an image conversion tool that transforms heterogeneous files (e.g., Excel and PDF) into visual representations. 
These tools make reported results directly viewable and convert heterogeneous reproduced artifacts into visual representations when needed, enabling reliable consistency checking.
% These tools make reported results and reproduced artifacts accessible in a unified multimodal form for consistency checking. 
The Report Agent uses a PDF rendering tool that converts structured Markdown reports into PDF files, enabling standardized and readable report generation.

\subsection{Key Design}

We present four designs that improve the reliability of \paperexec for end-to-end reproducibility assessment. 

\textbf{File Persistence.}
Many reproduction packages surface results only through GUIs or console prints, which disappear after execution and make evaluation unreliable. \paperexec therefore persists key information as files throughout the pipeline. Through \textit{enforced results saving}, it edits scripts when needed to explicitly save reproduced outputs (e.g., tables and figures) as files, preserving artifacts for downstream evaluation. Through \textit{log capture}, it records executions as log files to support failure diagnosis and process traceability. Through \textit{deliverable persistence}, it saves execution plans and summaries as files, making the process more transparent and supporting human review. This design improves evaluation reliability, enables inspection, and keeps the end-to-end process traceable.

\textbf{Context Control.}
% Reproducibility assessment requires interacting with many files, scripts, and verbose runtime traces, which can exceed an agent’s effective context capacity. \paperexec controls context growth through stage separation and role specialization, where each agent operates on a bounded subtask. We further regulate information flow with paginated file access, so agents retrieve only the necessary slices of large documents. For highly verbose execution outputs, the system retains representative segments (e.g., the head and tail of logs) to support debugging without flooding context. Finally, our lightweight editing interface supports content search and targeted replacement, which keeps iterative modifications compact and avoids accumulating redundant code snapshots.
Reproducibility assessment often involves many files, scripts, and verbose runtime traces, which can exceed an agent’s effective context capacity. We control context growth through multiple mechanisms throughout the pipeline. Through \textit{task decomposition}, each agent’s context is limited to the information needed for a bounded subtask. Through \textit{paginated file access}, agents selectively read only the needed portions of large documents without loading them in full. Through \textit{truncating execution feedback}, agents observe only the head and tail of verbose execution logs, which supports failure diagnosis without flooding the context. Through \textit{lightweight editing}, agents apply content search and targeted replacement, keeping iterative modifications compact and avoiding redundant code snapshots.

\textbf{Expert Prompting.}
% \paperexec designs agents to behave like domain experts by encoding task-specific skills into prompts. We provide each agent with procedural guidance that reflects how human reproducers plan executions, diagnose failures, and organize evidence. In addition, we employ few-shot prompting to teach effective tool-use and workflow patterns. For example, we include demonstrations on how to persist reproduction outputs as files, which helps agents execute and modify scripts more reliably, especially in specialized ecosystems such as R and Stata.
% Considering the challenges of these tasks, we design agents to behave like domain experts through prompts that encode task-specific skills. 
To handle the practical challenges of reproducibility assessment, we equip each agent with expert prompts that encode task-specific skills and decision rules.
Through \textit{procedural guidance}, we provide step-by-step instructions that mirror how human reproducers plan executions, diagnose failures, and organize evidence. Through \textit{few-shot prompting}, we teach effective tool use and workflow patterns. For example, we include demonstrations that show how to persist reproduced results as files, which helps agents execute and minimally modify scripts more reliably, particularly in specialized ecosystems such as R and Stata.

\textbf{Failure Recovery.}
% End-to-end reproducibility assessment involves many dependent steps, so a local failure can propagate and compromise downstream decisions. \paperexec reduces this risk by enabling recovery from partial failures across stages. When expected outputs are missing, downstream agents do not stop at the absence of a file; instead, they search for alternative evidence in intermediate artifacts and runtime logs and attempt to reconstruct the underlying reproduction result (e.g., by locating printed tables in logs). When execution feedback indicates unmet dependencies, later agents may also re-attempt essential setup steps. To make recovery safe and reversible, \paperexec applies non-intrusive patches by generating modified copies rather than overwriting source files. This design preserves the original repository state, keeps changes reviewable, and supports rollback when a patch introduces unintended side effects.
End-to-end reproducibility assessment involves many dependent steps, so a local failure can propagate and compromise downstream assessment. We therefore design multiple recovery mechanisms in \paperexec. Through \textit{robust continuation}, downstream agents remain robust to upstream failures and actively attempt recovery rather than collapsing. For example, when reproduced artifacts are missing, the Scoring Agent searches other artifacts and attempts to reconstruct the result, such as by locating tables printed in the logs. 
% The Execution Agent likewise repeats essential setup steps when it detects missing dependencies or an incomplete environment. 
Through \textit{non-intrusive modifications}, agents apply script changes by generating modified copies rather than overwriting source files. This preserves the original repository state, keeps changes reviewable, and supports rollback when an edit introduces unintended side effects.

\section{Experiments}

% \tx{How about list some research questions here as we have many results to discuss?}

% \begin{itemize}
%     \item \textbf{RQ1: }How does \paperexec perform compared to existing agents for social science reproducibility?
%     \item \textbf{RQ2: }Can \paperexec generalize across languages?
%     \item \textbf{RQ3: }How does each design in \paperexec  contributes to its overall performance?
%       \item \textbf{RQ4: }How does \paperexec  generalize to social science study beyond the benchmark?
%     \item
% \end{itemize}

\subsection{Experimental Setup}
\subsubsection{Benchmark.}

We evaluate on REPRO-Bench~\cite{hu2025repro}. It contains 112 instances from social science papers for computational reproducibility assessment. Each instance provides a paper, a reproduction package, and a list of major findings, and the system is required to generate a reproducibility score. REPRO-Bench uses a four-level rubric. Score 1 means major findings are irreproducible. Score 4 means major findings are fully reproducible. Scores 2 and 3 mean major findings are reproducible, but issues remain. Score 2 reflects minor inconsistencies or errors in the provided code that do not change major findings. Score 3 reflects minor differences in presentation, such as rounding or formatting.

\subsubsection{Our approach and Baselines.}
REPRO-Bench requires only a final reproducibility score. We therefore run \paperexec with the first three agents (Setup, Execution, and Scoring) in all experiments, and omit the Report Agent unless stated otherwise.
We compare with agent-based baselines from REPRO-Bench, including SWE-Agent~\cite{sweagent}, AutoGPT~\cite{AutoGPT}, CORE-Agent~\cite{siegel2024core}, and REPRO-Agent~\cite{hu2025repro}. SWE-Agent is a software engineering agent equipped with an agent-computer interface for repository interaction and debugging.
AutoGPT is a generalized agent that plans and iteratively interacts with tools.
CORE-Agent adapts AutoGPT with task guidance and structured output checks for reproducibility assessment.
REPRO-Agent further extends CORE-Agent with template-based planning, error-aware prompting, and a dummy-score fallback when assessment fails. Notably, such a fallback may obscure failures and yield unreliable scores. All approachs use GPT-4o~\cite{gpt4O} unless stated otherwise.

\subsubsection{Metrics.}
We report Accuracy, Applicability, and Executability. Accuracy is the fraction of instances where the predicted score matches the ground truth. Applicability is the fraction of instances that produce a valid output file in the required format and name. Executability is evaluated on instances with ground-truth scores in 2--4 and reports the fraction where the approach predicts a score in 2--4. It indicates whether the approach can run the reproduction package and obtain usable artifacts. For cost analysis, we report the average API cost per instance for each approach. More details about the experimental setup can be found in Appendix~\ref{app:experiment_setup}.

\subsection{Experimental Results and Analysis}
\subsubsection{Main Results.}
\begin{table}
\caption{Performance and costs of different approachs on REPRO-Bench. “–” denotes unavailable metrics that are not reported in the original paper and lack open-sourced code.}
\centering
\resizebox{\columnwidth}{!}{%
\begin{tabular}{lcccc}
\toprule
Approach & \% Accuracy & \% Applicability & \% Executability & \$ Cost \\
\midrule
SWE-Agent      & 10.7 & 19.6 & 16.3 & 1.20 \\
AutoGPT        & 20.5 & 60.7 & 45.7 & 2.03  \\
CORE-Agent     & 21.4 & 46.4 & 39.1 & 2.00  \\
REPRO-Agent     & 36.6 & 92.9 & - & -  \\
\paperexec      & \textbf{44.6} & \textbf{100.0} & \textbf{65.2} & 1.93 \\
\bottomrule
\end{tabular}
}
\label{tab:model_performance}
\end{table}

% Table~\ref{tab:model_performance} summarizes performance on REPRO-Bench. 
% \paperexec consistently outperforms all baselines across accuracy, applicability, and executability. 
% Compared with the strongest baseline REPRO-Agent, \paperexec improves accuracy by 36.8\% (relative), indicating more reliable end-to-end reproducibility assessment under the four-level rubric.
% In contrast, AutoGPT and CORE-Agent remain close to random guessing among four labels, and SWE-Agent performs substantially worse, indicating that software engineering agents do not directly transfer to social science reproducibility assessment. \paperexec also achieves perfect applicability, while several baselines frequently fail to produce valid outputs. Our stage separation drives this improvement. The execution stage collects reproduction artifacts, and the Scoring Agent operates with a fresh context and assigns the final score even when earlier steps only partially succeed.
% REPRO-Agent reports high applicability, but its dummy-score fallback can inflate both applicability and accuracy when the agent fails to complete the assessment.
% Finally, \paperexec yields the highest executability, suggesting that it more reliably completes executions and produces usable reproduction artifacts.

Table~\ref{tab:model_performance} reports results on REPRO-Bench. \paperexec achieves the best performance on all reported metrics with 44.6\% accuracy, 100.0\% applicability, and 65.2\% executability, and its cost is comparable to prior approaches at \$1.93 per instance. REPRO-Agent is the strongest baseline on accuracy at 36.6\%, and \paperexec improves it by 8.0 points, which is a 21.9\% relative gain. This indicates more reliable end-to-end scoring under the four-level rubric. For applicability, \paperexec is the only approach that produces valid outputs for all instances, while REPRO-Agent reaches 92.9\% and other baselines fail more often. This suggests our approach is more robust to partial failures and still returns a well-formed score file. For executability, \paperexec outperforms AutoGPT at 45.7\% and CORE-Agent at 39.1\%. This suggests it successfully executes and collects reproduced artifacts more often before assigning a score.

\subsubsection{Score Distribution Analysis.}
\label{Score}
\begin{figure}[t]
  \includegraphics[width=\columnwidth]{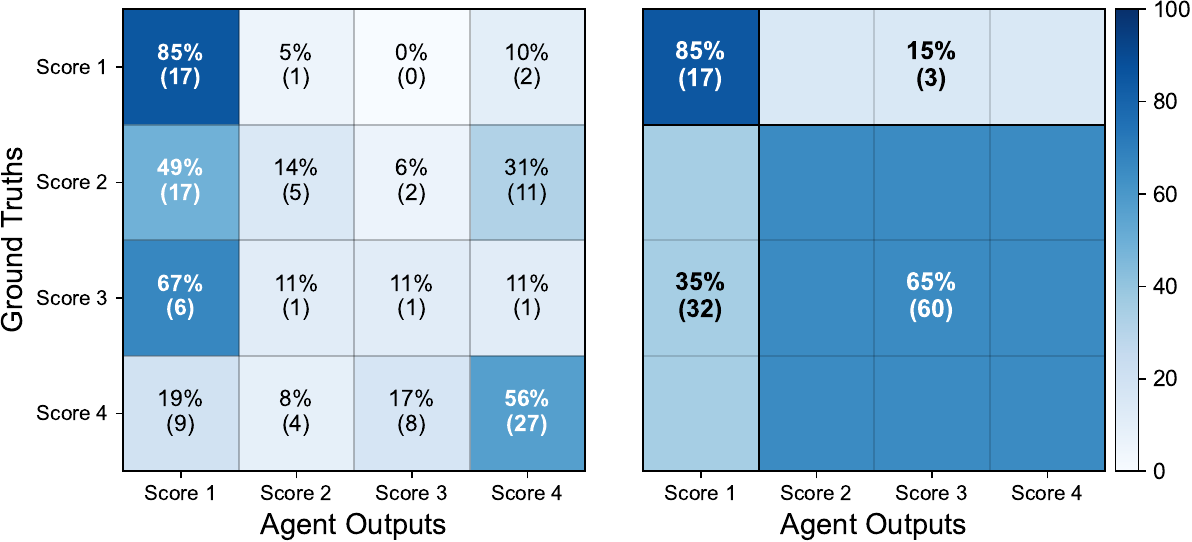}
  \caption{Score-level confusion heatmaps for \paperexec. Left: normalized distribution over scores 1--4. Right: scores 2--4 merged, consistent with the executability metric.}
  \label{fig:score}
\end{figure}

% Figure~\ref{fig:score} shows how \paperexec’s predictions distribute across ground-truth scores. \paperexec performs best at the two ends of the rubric. It distinguishes fully irreproducible instances and fully reproducible instances more reliably than the intermediate instances. Most errors concentrate in the intermediate scores. Instances with ground-truth scores of 2 or 3 require detecting subtle issues, such as minor code-level inconsistencies or presentation-level mismatches, which are easy to miss without careful alignment between artifacts and the paper. As a result, \paperexec sometimes assigns a score of 4 to these instances and misses the underlying issues captured by the benchmark. \paperexec also assigns a score of 1 to many such instances, which may indicate that these instances are harder to execute or harder to align with the reported results. The right panel merges scores 2--4 as executable, aligning with our executability metric because these instances typically execute and produce usable artifacts even when minor issues remain. Under this aggregation, \paperexec separates non-executable and executable instances more cleanly.

Figure~\ref{fig:score} shows the distribution of \paperexec predictions across ground-truth scores. Performance is strongest at the two ends of the rubric. For score 1, \paperexec predicts 1 for 85\% of instances, while for score 4 it predicts 4 for 56\% of instances. Errors concentrate in the intermediate labels. For score 2, only 14\% are predicted as 2, and many are predicted as 1 or 4. For score 3, most instances are predicted as 1, and only 11\% are predicted as 3. This pattern suggests that distinguishing scores 2 and 3 requires detecting subtle implementation or presentation issues that are often missed. The right panel groups scores 2–4 as an indicator of successful execution, matching our executability metric. Under this view, \paperexec assigns scores 2–4 to 65\% of instances with a true score in 2–4. This indicates a higher rate of successful execution with usable reproduced artifacts. We also observe a small number of instances with a true score of 1 that still execute successfully, where \paperexec may overlook severe mismatches between the code implementation and the paper.

\subsubsection{Failure Analysis.}
\label{failure_analysis}
\begin{figure}[t]
  \includegraphics[width=\columnwidth]{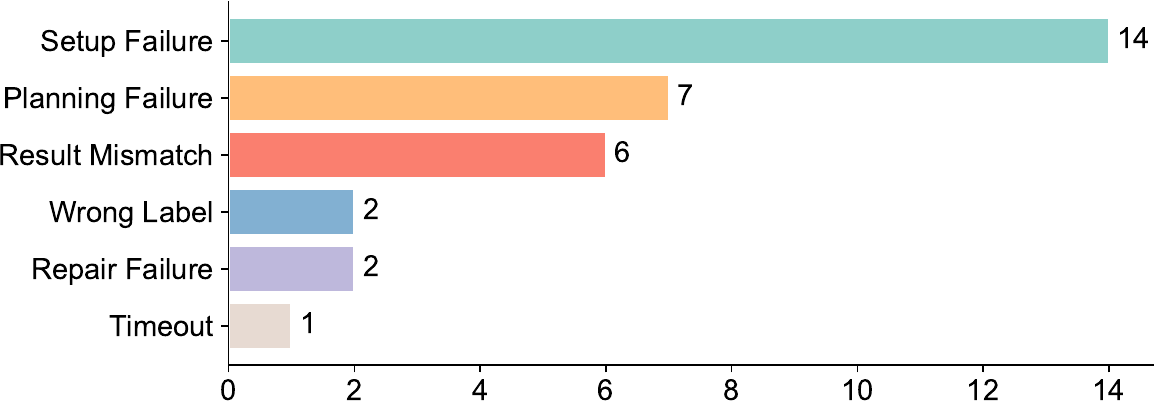}
  \caption{Distribution of failure types for cases where \paperexec predicts score 1, but the true score is in 2--4.}
  \label{fig:failures}
\end{figure}

% Figure~\ref{fig:score} categorizes failures for instances where \paperexec predicts score 1 while the ground-truth score is in 2--4. Setup failures are the most common. They occur when the system cannot build a runnable environment due to missing dependencies or incomplete setup instructions. Planning failures form the second largest group. They arise when the system selects the wrong entry script or executes scripts in an incorrect order, which prevents it from producing the required artifacts. We also observe cases where execution completes, but the results do not match the benchmark expectation. In these instances, the system successfully executes the scripts, but the generated artifacts cannot be mapped to the intended reproduction items, or the relevant artifacts are not saved. In addition, we identify a small number of benchmark label errors \tx{may need to specify}. By cross-checking the original benchmark with reproduction reports and repositories, we find two instances that miss required data or dependencies but are labeled as executable in the ground truth\tx{may need to specify}. The remaining failure types are less frequent, including repair failures and timeouts. Beyond execution-related errors, \paperexec can also make incorrect consistency judgments. It sometimes assigns a high score when the implementation substantially deviates from the paper. It can also miss minor code-level issues associated with score 2 and subtle reporting-level discrepancies associated with score 3.

Figure~\ref{fig:failures} summarizes failures for instances where \paperexec predicts score 1 while the true score is in 2--4. Setup failures are most common with 14 cases, usually due to missing dependencies, incompatible versions, or unclear setup instructions. Planning failures follow with 7 cases, where the system selects an incorrect entry point or runs scripts in an order that prevents artifact generation. Result mismatches account for 6 cases. In these cases, execution completes, but the produced files cannot be mapped to the intended reproduction items, or the required outputs are not saved. In particular, we find 2 failure cases that may reflect benchmark annotation issues. The corresponding instances miss required data or dependencies but are labeled with scores in 2--4. The remaining failures are rare, including 2 repair failures and 1 timeout.
Beyond failures in execution and artifact collection, \paperexec can still make scoring mistakes. As discussed in Section~\ref{Score}, it occasionally assigns a high score when the implementation substantially deviates from the paper, and it can miss minor issues that separate scores 2 and 3.

\subsubsection{Language-wise Performance Analysis.}
\begin{table}[t]
\caption{Performance under different programming language settings.}
\centering
\begin{tabular}{lccc}
\toprule
Language Setting & Instances & \% Accuracy & \% Executability \\
\midrule
R-only & 25 & 56.0 & 73.7 \\
Stata-only & 63 & 38.1 & 61.4 \\
Multi-language & 17 & 47.1 & 76.9 \\
\bottomrule
\end{tabular}
\label{tab:lang_setting_perf}
\end{table}

% Table~\ref{tab:lang_setting_perf} reports performance under different language settings. \paperexec performs best on R-only packages and worst on Stata-only packages. R repositories often provide clearer script-level entry points, which simplifies planning and execution. 
% Stata workflows typically require more careful environment configuration and project-specific conventions, which increases execution friction. 
% In addition, R is more widely used and publicly documented than Stata, so LMs likely have stronger prior knowledge of R syntax and common libraries.
% Interestingly, multi-language packages achieve strong executability and competitive accuracy, possibly because coordinating multiple toolchains encourages more structured repositories and more explicit documentation.

Table~\ref{tab:lang_setting_perf} reports performance under different language settings. \paperexec achieves the highest accuracy on R-only packages at 56.0\% and the lowest on Stata-only packages at 38.1\%. Executability shows a similar trend, with 73.7\% for R-only and 61.4\% for Stata-only. One likely factor is that many R repositories expose clearer script entry points, which simplifies planning and execution. Stata workflows more often rely on environment setup and conventions specific to a project, which can increase execution friction.
Multi-language packages reach strong executability at 76.9\% with competitive accuracy at 47.1\%. This may reflect better repository structure and more explicit documentation when multiple toolchains must be coordinated.

\subsection{Additional Analysis}
\subsubsection{Ablation Study.}

\begin{table}[t]
\caption{Ablation results on specialized agents and task-specific tools. Results are computed on a random subset of 30 instances.}
\centering
\begin{tabular}{lcc}
\toprule
Variant & \% Accuracy & \% Executability \\
\midrule
\paperexec (full) & \textbf{43.3} & \textbf{70.8} \\
\midrule

\multicolumn{3}{c}{\textbf{Agent Ablations}} \\
\midrule
Merge Setup and Execution & 33.3 & 58.3 \\
Merge Execution and Scoring & 30.0 & 45.8 \\
\midrule

\multicolumn{3}{c}{\textbf{Tool Ablations}} \\
\midrule
Execution: w/o editing tool & 36.7 & 41.7 \\
Scoring: w/o extraction tool & 40.0 & 66.7 \\
Scoring: w/o viewing tool & 26.7 & 45.8 \\
\bottomrule
\end{tabular}
\label{tab:ablation}
\end{table}

To study the effects of separating agent responsibilities and using task-specific tools, we conduct ablations, as shown in Table~\ref{tab:ablation}. All variants reduce performance compared with the full system, which supports the value of agent separation and task-specific tools.

\textbf{Agent ablations.}
\textit{Merge Setup and Execution} assigns the Execution Agent to first configure the environment and then execute the repository. This reduces accuracy from 43.3\% to 33.3\% and executability from 70.8\% to 58.3\%.
\textit{Merge Execution and Scoring} assigns the Execution Agent to execute the repository and then generate the final reproducibility score. This further reduces accuracy to 30.0\% and executability to 45.8\%.
Overall, merging agent responsibilities consistently degrades both metrics, highlighting the value of keeping setup, execution, and scoring in separate agents.

\textbf{Tool ablations.}
\textit{Execution without the editing tool} removes the editing tool from the Execution Agent, so code changes rely on rewriting files. This drops accuracy to 36.7\% and executability to 41.7\%.
\textit{Scoring without the extraction tool} removes the extraction tool from the Scoring Agent. The agent must inspect the full paper images to locate relevant figures and tables instead of directly retrieving the target screenshots. This drops accuracy to 40.0\% and executability to 66.7\%.
\textit{Scoring without the viewing tool} removes the viewing tool from the Scoring Agent and replaces it with a query-based vision language model tool for indirect image access. This drops accuracy to 26.7\% and executability to 45.8\%. This variant yields the largest accuracy drop, suggesting that indirect access to figures and tables may be insufficient for reliable visual matching.
Overall, removing any specific tool reduces performance, supporting the contribution of task-specific tools.

\subsubsection{Model Impact.}

We study how the backbone LM affects each agent by swapping GPT-4o with GPT-4o-mini in one agent at a time while keeping the rest unchanged. Using weaker backbones consistently reduces performance, with the Execution Agent most sensitive and the Setup Agent relatively tolerant. The results highlight a practical advantage of our multi-agent design. It enables allocating different model strengths to different agents based on task needs, reducing cost without changing the overall pipeline. Detailed results are reported in Appendix~\ref{app:model_impact}.

\subsubsection{Generalization Analysis.}

To examine whether \paperexec generalizes beyond REPRO-Bench, we conduct a case study on a Nature cover article with an open replication package~\cite{ratledge2022using}. We target a multi-panel figure with two panels that integrates three causal estimators. Reproducing the figure involves six R scripts with strong cross-script dependencies, so they must be executed in a specific order. The scripts contain 2,364 lines in total, including one 989-line script, which makes debugging and reliable end-to-end execution challenging. The package also assumes environment-specific paths and recommends running in RStudio, which makes it harder to execute in an automated setting. Despite these challenges, \paperexec completes the reproducibility assessment. In the execution stage, it identifies the six scripts and required datasets, infers the correct execution order, fixes runtime issues, and saves the reproduced outputs as two PDF files. In the evaluation stage, it compares the generated PDFs with the corresponding figure in the paper, assigns a reproducibility score, and produces a structured report. Appendix~\ref{app:case} includes the full process.

We also compare \paperexec with a human expert on the same task. The expert takes about one hour to assess reproducibility, while \paperexec completes the assessment in about 15 minutes.

\textbf{Implications.}
This study suggests that \paperexec can support computational reproducibility assessment for complex, publication-grade analytical workflows. Integrating \paperexec into the research process can reduce assessment time and make pre-publication checks more feasible. Moreover, \paperexec produces traceable execution artifacts and structured reports, facilitating human review and independent auditing.

\section{Rethink Repro-Bench}
\label{sec:reprobench_s}

% This section revisits ReproBench and presents a refined benchmark, \textbf{ReproBench-S}, together with additional evaluation and analysis. 
% In practice, we observed that a non-trivial portion of errors on ReproBench were caused by benchmark issues rather than model limitations. To enable more reliable and diagnostic evaluation, we conduct a full audit of all instances, correct inconsistent or invalid annotations, and further stratify instances by execution difficulty.

\subsection{REPRO-Bench-S}
\label{subsec:reprobench_s_construction}

% As our failure analysis reveals, REPRO-Bench contains instances with incorrect ground-truth annotations, which can misattribute execution failures to the system and introduce noise into evaluation. We therefore conduct a comprehensive audit of all 112 instances by cross-checking the benchmark annotations against the original public reproduction reports and the associated repositories. Based on the verified instances, we further stratify the benchmark by execution difficulty, resulting in REPRO-Bench-S.

As discussed in Section~\ref{failure_analysis}, REPRO-Bench contains instances with incorrect annotations, which introduce noise into evaluation. We therefore audit all 112 instances using public reproduction reports and repositories. We also stratify the verified instances by execution difficulty to enable more diagnostic analysis and form REPRO-Bench-S.

\subsubsection{Instance correction.}
Three authors with at least seven years of programming experience independently review all candidate issues and resolve disagreements through discussion. We identify 13 problematic instances and apply three types of corrections. We remove reproduction items that are not supported by the package, adjust labels when reproduction reports discuss issues outside the specified items so that labels reflect the specified items only, and fix incorrect labels based on evidence from the reports and repositories. Correction details are provided in Appendix~\ref{app:repro_bench_correction}.
% Four authors with at least seven years of programming experience independently reviewed all candidate issues and resolved disagreements through group discussion. We identify 11 problematic instances and apply three types of corrections:
% \begin{itemize}
% \item[(i)] \textbf{Item removal.} We remove extra reproduction items that are not supported by the reproduction package. This includes items that cannot be linked to any script, intermediate file, or derivable artifact in the repository.

% \item[(ii)] \textbf{Scope alignment.} In some instances, the issues documented in the original reproduction report concern results outside the benchmark-specified reproduction items. We revise the ground-truth labels to reflect the reproducibility status of the specified items only.

% \item[(iii)] \textbf{Label correction.} We fix factual misannotations in the ground truth. This includes instances marked as reproducible despite missing required data or dependencies, and instances whose reproducibility score conflicts with the evidence in the reproduction report and repository.
% \end{itemize}

\subsubsection{Difficulty stratification.}
To support more diagnostic evaluation, we stratify instances by execution difficulty. Our difficulty criteria capture the clarity of execution guidance, the amount of required intervention, and the ease of identifying outputs. Concretely, we consider whether the documentation specifies a clear entry point and run order, how many files require modification, whether outputs must be explicitly saved, and how directly produced artifacts map to the benchmark reproduction items. We assign each instance to one of three difficulty levels, denoted as Level-1, Level-2, and Level-3, corresponding to easy, medium, and hard. Three authors label all instances independently, and we use majority vote and discuss any disagreements to reach a final decision. Stratification details are provided in Appendix~\ref{app:repro_bench_stratification}.

% Instances vary substantially in how difficult they are to execute and obtain the required outputs for the reproduction items. We therefore annotate each instance with one of three execution difficulty levels:
% \begin{itemize}
% \item \textbf{Level-1 (easy).} The repository documentation clearly specifies how to execute the reproduction package to obtain the outputs for each reproduction item. Execution requires editing at most two files, and the edits are limited to minor path fixes. After execution, the artifact(s) for each reproduction item can be identified directly.

% \item \textbf{Level-2 (medium).} Instances that do not meet the Level-1 criteria but still provide relatively clear execution guidance. Execution requires editing at most four files. After execution, the artifact(s) corresponding to each reproduction item remain easy to identify.

% \item \textbf{Level-3 (hard).} Instances that do not meet the Level-1 or Level-2 criteria. This typically involves unclear execution instructions, more than four files requiring modification, or hard output identification, where execution generates many candidate artifacts, and the system must infer which one(s) correspond to each reproduction item.
% \end{itemize}
% Four authors labeled all instances independently. We take the majority vote as the final label and resolve ties through discussion to reach consensus.

\subsubsection{Dataset characteristics.}
\begin{table}[t]
\caption{Score distribution across difficulty levels on REPRO-Bench-S.}
\label{tab:level_score}
\centering
\begin{tabular}{lrrrrr}
\toprule
Level & Score 1 & Score 2 & Score 3 & Score 4 & Total \\
\midrule
Level-1 & 6  & 4  & 1  & 17  & 28 \\
Level-2 & 6  & 8  & 2 & 20  & 36 \\
Level-3 & 10  & 18  & 6  & 14 & 48 \\
\midrule
Total & 22 & 30 & 9 & 51 & 112 \\
\bottomrule
\end{tabular}
\end{table}

Table~\ref{tab:level_score} reports the composition of REPRO-Bench-S by execution difficulty and score.
Overall, score 4 is the largest group, while score 3 is rare.
The score distribution also shifts with difficulty. Level-1 and Level-2 are dominated by score 4, whereas Level-3 contains fewer score 4 instances and more lower-score instances, especially 2 and 3.
This indicates that harder instances are associated with more implementation or reporting discrepancies. The difficulty labels provide an additional axis for analysis and enable more diagnostic evaluation.
% This pattern suggests a correlation between execution difficulty and score, with Level-3 instances more likely to receive intermediate (2--3) or lower scores.
% The difficulty labels provide an additional axis for analysis and support more diagnostic evaluation across increasingly challenging execution settings.

% \begin{figure}[t]
%     \centering
%     % left: score distribution; right: difficulty distribution
%     \includegraphics[width=\linewidth]{figs/reprobench_s_stats.pdf}
%     \caption{ReproBench-S statistics: (left) reproducibility score distribution; (right) difficulty-level distribution.}
%     \label{fig:reprobench_s_stats}
% \end{figure}

\subsection{Performance on REPRO-Bench-S}
\label{subsec:reprobench_s_results}

\begin{table}
\caption{Performance of different approaches on REPRO-Bench-S.}
\label{tab:model_performance_reprobench_s}
\centering
\begin{tabular}{lcccc}
\toprule
Approach & \% Accuracy & \% Applicability & \% Executability\\
\midrule
SWE-Agent      & 11.6 & 21.4 & 16.7 \\
AutoGPT        & 21.4 & 63.4 & 48.9  \\
CORE-Agent     & 23.2 & 49.1 & 41.1  \\
\paperexec      & \textbf{50.9} & \textbf{100.0} & \textbf{66.7} \\
\bottomrule
\end{tabular}
\end{table}

\begin{table}[t]
\caption{Performance by difficulty level on REPRO-Bench-S.}
\label{tab:performance_reprobench_s}
\centering
\begin{tabular}{lccc}
\toprule
Level & Instance Number & \% Accuracy & \% Executability \\
\midrule
Level-1 & 28 & 75.0 & 86.4 \\
Level-2 & 36 & 55.6 & 76.7 \\
Level-3 & 48 & 33.3 & 47.4 \\
\midrule
Overall & 112 & 50.9 & 66.7 \\
\bottomrule
\end{tabular}
\end{table}

We evaluate \paperexec on REPRO-Bench-S and compare it with prior agent-based baselines. We omit REPRO-Agent because it is not publicly available and its paper does not report the score distribution needed to estimate its performance. Table~\ref{tab:model_performance_reprobench_s} reports overall results on REPRO-Bench-S. 
\paperexec achieves the best accuracy at 50.9\% and the best executability at 66.7\%, while maintaining perfect applicability. REPRO-Bench-S removes a portion of noise caused by inconsistencies in the original benchmark, and the improved scores therefore better reflect true system capability.

Table~\ref{tab:performance_reprobench_s} summarizes \paperexec’s performance across difficulty levels. Accuracy decreases from 75.0\% (Level-1) to 33.3\% (Level-3), and executability similarly drops from 86.4\% to 47.4\%. This trend aligns with our stratification, since harder instances are more difficult to execute, which in turn makes reproducibility assessment more challenging.
The results also indicate that execution robustness remains the primary bottleneck for harder instances.

\section{Related Work}
\textbf{Computational Reproducibility Assessment.} 
Recent work explores AI agents for computational reproducibility assessment and has introduced several benchmarks and agent systems. Among these benchmarks, SciCode~\cite{tian2024scicode} casts reproducibility tasks as code generation problems derived from a paper’s main findings. CORE-Bench~\cite{siegel2024core} builds multi-step reproduction tasks from verified CodeOcean repositories and evaluates agents through questions based on execution results. REPRO-Bench~\cite{hu2025repro} uses a four-level rubric that considers both result consistency and implementation consistency, evaluating whether agents can accurately assess reproducibility for social science papers. 
Based on these benchmarks, several agent systems have been proposed. AutoGPT~\cite{AutoGPT} has been adapted to reproducibility assessment by equipping it with execution and vision-language tools for analyzing figures. CORE-Agent~\cite{siegel2024core} further specializes AutoGPT with task-specific prompting and structured output checking tailored to CORE-Bench. Building on CORE-Agent, REPRO-Agent~\cite{hu2025repro} introduces template-based planning and error-aware prompting to mitigate common failure patterns observed in reproducibility tasks. Despite these advances, existing agent-based systems remain limited for social science reproducibility. They are limited by insufficient context and inadequate tool support across the assessment process. In contrast, \paperexec decomposes assessment into subtasks handled by specialized agents, equipped with task-specific tools and expert prompts to better handle these cases.

\textbf{LM-Based Agents for Software Engineering.}
% Assessing computational reproducibility involves environment configuration, code execution, error diagnosis, and result inspection, making it closely related to complex software engineering workflows. Accordingly, a growing body of work has explored LLM-based agents for software engineering tasks. Repo2Run focuses on automated environment setup by iteratively constructing executable Docker environments for arbitrary repositories. Other agents target software maintenance and debugging: SWE-Agent designs a carefully engineered agent–computer interface that enables LLMs to navigate repositories and iteratively resolve issues, while AutoCodeRover and CodeR combine LLMs with code search, program structure analysis, and multi-agent coordination for issue resolution. Beyond execution and debugging, agent-based approaches have also been proposed for code evaluation. CodeVisionary introduces a two-stage framework that distills multi-dimensional contextual information and performs fine-grained code assessment through coordinated agent judgments. While these systems demonstrate strong capabilities in handling complex code-centric tasks, they are not designed for reproducibility assessment as a judgment problem, and do not explicitly address result–paper alignment or consistency evaluation required in social science reproducibility assessment.
Computational reproducibility assessment encompasses many software engineering tasks, including environment setup, debugging, and code evaluation. Accordingly, we review LM-based agents along with these software engineering tasks. Recent LM-based agent frameworks combine LMs with planning, memory, perception, and tool use, which allows agents to interact with real repositories and execution environments over multiple steps~\cite{DBLP:journals/corr/abs-2309-07864,plan3,memory3,memory4,perception4}. For environment setup, agents iteratively configure dependencies and build runnable environments for arbitrary repositories~\cite{repo2run,executionagent}. For debugging, agents navigate codebases and leverage execution feedback to iteratively resolve issues~\cite{sweagent,openhands,autocoderover,coder,codev}. For code evaluation and review, agents use multi-stage designs to gather context perform fine-grained assessment through coordinated judging~\cite{codevisionary}. Beyond these three tasks, LM-based agents have also been applied to other software engineering tasks, including requirements engineering~\cite{DBLP:journals/corr/abs-2405-03256, DBLP:journals/corr/abs-2310-13976}, code generation~\cite{DBLP:journals/corr/abs-2312-13010, DBLP:journals/corr/abs-2404-02183}, unit testing~\cite{DBLP:journals/corr/abs-2305-04207}, system testing~\cite{DBLP:journals/corr/abs-2311-08649, DBLP:journals/corr/abs-2308-06782}. Despite these advances, most software engineering agents are designed for repository-level development and maintenance. Reproducibility assessment instead requires assessing consistency between the paper and the repository, including both the implementation and the reported result presentation, which remains challenging for existing agents~\cite{sweagent,codevisionary}. Our \paperexec fills this gap with explicit paper–code alignment and consistency evaluation, enabling precise reproducibility assessment.

\section{Conclusion}
We presented \paperexec, a novel two-stage, multi-agent approach for automated computational reproducibility assessment of social science papers. In the execution stage, agents execute the reproduction package and capture reproduced results as explicit artifacts. In the evaluation stage, agents evaluate reproducibility using explicit evidence. 
% Experiments on REPRO-Bench demonstrate strong performance, and we further refine the benchmark by introducing REPRO-Bench-S, a difficulty-stratified version to enable more diagnostic evaluation and support future research. Looking ahead, we plan to extend \paperexec beyond social science to additional scientific domains.
Experiments on REPRO-Bench demonstrate strong performance. We also refine the benchmark by correcting identified issues and introducing REPRO-Bench-S, a difficulty-stratified variant that enables more diagnostic evaluation and supports future research. Looking ahead, we plan to extend \paperexec beyond social science to additional scientific domains.

\section*{Limitations and Ethical Considerations}
Although \paperexec demonstrates strong performance, it still has limitations. First, \paperexec can struggle with minor paper--code mismatches and small code errors. While we incorporate code-quality checks, such issues can still be easy to miss. Second, when the repository environment is complex or the instructions are unclear, \paperexec may be unable to execute the repository. A specified Setup Agent and failure recovery mechanisms mitigate this, but execution failures can still occur.

% This work automates reproducibility assessment for social science papers. All experiments use the public REPRO-Bench and involve no human subjects, so no ethical concerns are implicated.

This work uses the public REPRO-Bench dataset, involves no human subjects, and does not use any sensitive personal data. We include safeguards to constrain agent actions and reduce safety risks, including prompt constraints and optional Docker isolation.
% In particular, \paperexec should be used as decision support, and final judgments should be made by human reviewers.
% We recommend using \paperexec as decision support rather than a sole basis for decisions, with human review of the produced evidence and artifacts.

\section*{GenAI Disclosure}
\textbf{Research usage.} This work uses large models as core components of our agent-based approach for automated reproducibility assessment. In our implementation, we instantiate agents with GPT-4o and GPT-4o-mini to assess reproducibility for target papers. We include safeguards, including prompt constraints and optional container isolation, to constrain agent actions during execution.

\textbf{Language polishing.} We use ChatGPT for language polishing of the manuscript only. All AI-assisted text is reviewed and edited by the authors, who take full responsibility for the final paper.


% \begin{acks}
% \end{acks}

\bibliographystyle{ACM-Reference-Format}
\bibliography{custom}

\appendix

\section{Additional Experimental Details}
\subsection{Additional Experimental Setup}
\label{app:experiment_setup}
We follow the standard REPRO-Bench setup. Each agent starts in a workspace that contains the paper PDF and a reproduction package directory. The environment preinstalls necessary software (e.g., Stata, MATLAB, R) and common dependency libraries, and agents can execute command-line operations with feedback from standard output and standard error streams. Considering the randomness of LLM generation, we run each experiment twice to determine the final result. If both runs are incorrect, we use the result from the first run. We terminate the agents if they incur
API costs of over \$4 per task. For the baseline results, we adopted the results reported in their papers.

\subsection{Backbone Model Impact Details}
\label{app:model_impact}
To understand how backbone LM affects each agent, we replace GPT-4o with GPT-4o-mini in one agent at a time and keep the other agents unchanged. As shown in Table~\ref{tab:model_impact}, using GPT-4o-mini in any single agent reduces both accuracy and executability compared with using GPT-4o for all agents. The largest impact comes from swapping the Execution Agent, where executability drops from 72.0\% to 48.0\% and accuracy drops to 40.0\%, while also yielding the largest cost reduction from \$1.67 to \$0.81. Swapping the Setup Agent has the smallest effect, with accuracy at 43.3\% and executability at 68.0\%, and lowers cost to \$1.36. Swapping the Scoring Agent mainly affects accuracy, which drops to 40.0\%, while executability remains relatively high at 64.0\% and cost decreases slightly to \$1.52. These trends indicate that the Execution Agent is most sensitive to model strength, while the Setup Agent is comparatively tolerant, and swapping the Scoring Agent primarily hurts accuracy, consistent with evaluation requiring careful evidence alignment. 

We further examine the cost breakdown when all agents use GPT-4o. Execution accounts for the majority of the total cost at \$1.035 out of \$1.67, while Setup and Scoring contribute \$0.326 and \$0.312, respectively. This suggests that model choices for the Execution Agent dominate end-to-end cost.

More broadly, the results highlight a practical advantage of our multi-agent design. It enables allocating different model strengths to different agents based on task needs, reducing cost without changing the overall pipeline.

\begin{table}[]
\caption{Impact of the backbone LM used by each agent. Results are computed on a random subset of 30 instances.}
\label{tab:model_impact}
\centering
\resizebox{\columnwidth}{!}{%
\begin{tabular}{lccc}
\toprule
LMs (Setup / Execution / Scoring)  & \% Accuracy & \% Executability & \$ Cost\\
\midrule
GPT-4o / GPT-4o / GPT-4o & 46.7 & 72.0 & 1.67\\
GPT-4o-mini / GPT-4o / GPT-4o & 43.3 & 68.0 & 1.36\\
GPT-4o / GPT-4o-mini / GPT-4o & 40.0 & 48.0 & 0.81\\
GPT-4o / GPT-4o / GPT-4o-mini & 40.0 & 64.0 & 1.52\\
\bottomrule
\end{tabular}
}
\end{table}

\subsection{Case Study Details}
\label{app:case}

\begin{figure}[H]
  \includegraphics[width=\columnwidth]{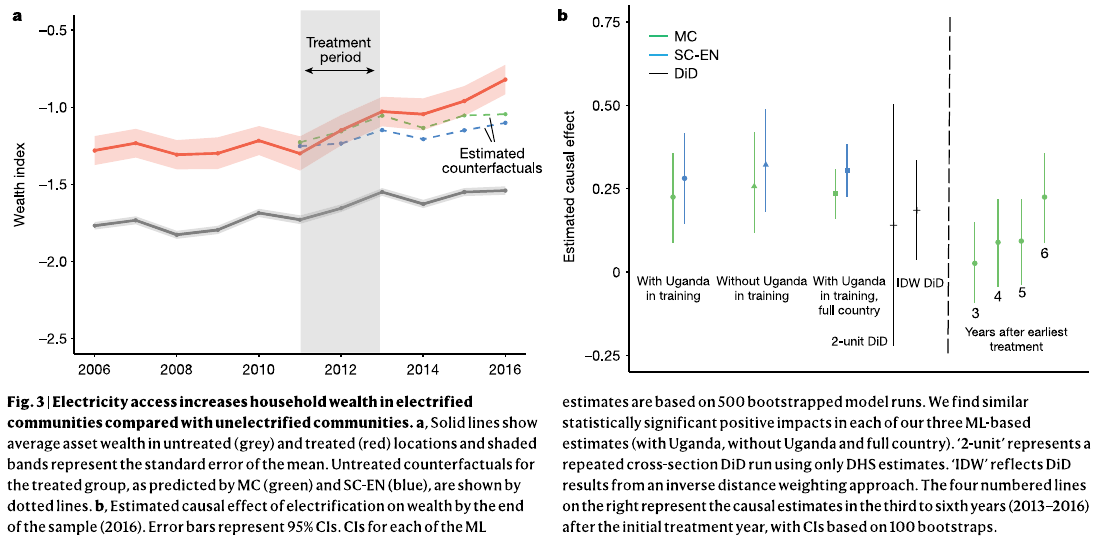}
  \caption{Target multi-panel figure selected for reproduction in the case study.}
  \label{fig:item}
\end{figure}

\begin{figure}[H]
  \centering

  \begin{subfigure}[t]{0.49\columnwidth}
    \centering
    \includegraphics[width=\linewidth]{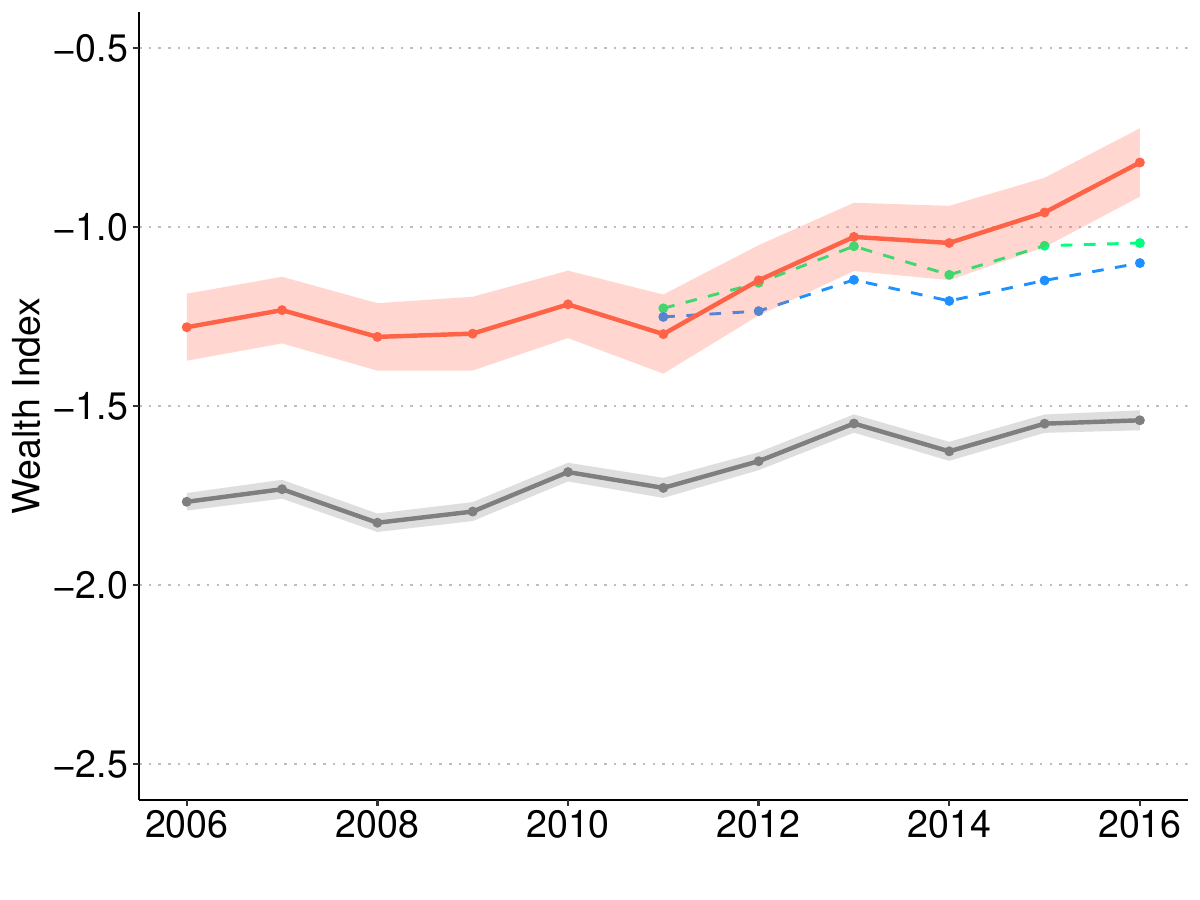}
    \caption{Reproduced left panel.}
    \label{fig:item-a}
  \end{subfigure}
  \hfill
  \begin{subfigure}[t]{0.49\columnwidth}
    \centering
    \includegraphics[width=\linewidth]{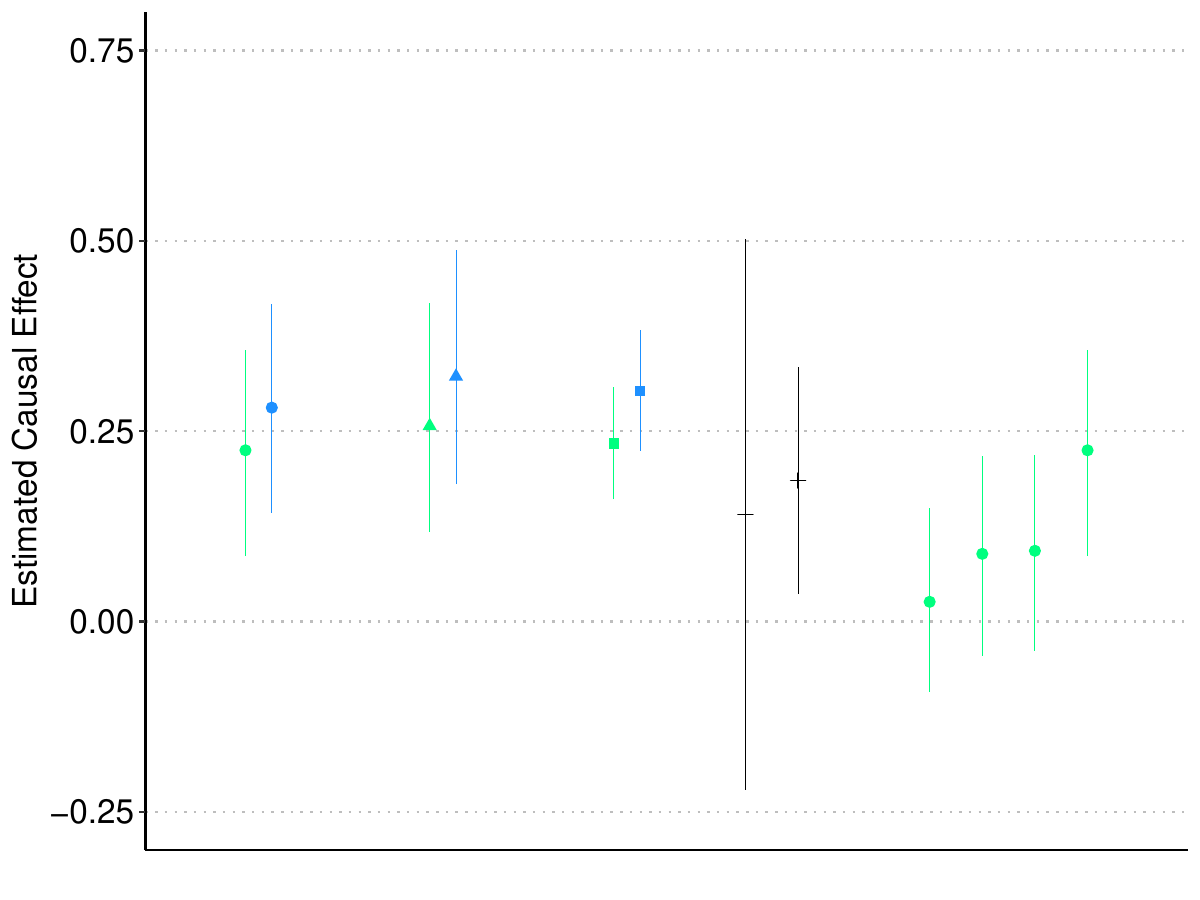}
    \caption{Reproduced right panel.}
    \label{fig:item-b}
  \end{subfigure}

  \caption{Reproduced two-panel figure from the case study.}
  \label{fig:repro_item}
\end{figure}

We assess reproducibility by reproducing the target multi-panel Figure~\ref{fig:item} from the Nature case study. The target figure integrates results from three causal estimators, including difference-in-differences (DiD), matrix completion (MC), and synthetic control with elastic net (SC-EN). This requires running multiple upstream scripts to compute intermediate results before the final plotting script can assemble the multi-panel figure, which creates strong cross-script coupling and a strict execution order. Since the required R package MPanel is hosted on GitHub, we install it in advance, and because running 500 model iterations is too time consuming, we reduce the runs to 2. \paperexec completes the end-to-end pipeline, including environment setup, script execution, output scoring, and report generation. Figure~\ref{fig:repro_item} shows the reproduced two-panel outputs generated by \paperexec, which is visually nearly identical to Figure~\ref{fig:item}. Figure~\ref{fig:setup} presents the reproduction plan produced by the Setup Agent, which enumerates the required files and specifies the strict execution order. Figure~\ref{fig:execution} provides the execution summary produced by the Execution Agent, including the identified scripts, necessary path and output adjustments, and generated figure artifacts. Figure~\ref{fig:scoring} shows the scoring summary produced by the Scoring Agent, which compares the reproduced outputs against the original figure and assigns a reproducibility score with a brief evaluation. Figure~\ref{fig:report} presents the reproduction report produced by the Report Agent, which consolidates the end-to-end process and results into a structured record.

\begin{figure*}[htbp]
\centering
\begin{tcolorbox}[
  colback=white,
  colframe=black,
  title=Reproduction Plan,
  % breakable,
  width=\linewidth,
  boxsep=1mm,
  left=-1mm,right=-1mm,top=-1mm,bottom=-1mm
]
\begin{tcblisting}{
  listing only,
  width=\linewidth,
  boxsep=0mm,
  left=0mm,right=0mm,top=0mm,bottom=0mm,
  sharp corners,
  arc=0mm,
  listing options={
    basicstyle=\ttfamily\footnotesize,
    breaklines=true,
    columns=fullflexible,
    keepspaces=true
  }
}
{
  "Figure 3": {
    "related_files": [
      "/nature/replication_package/scripts/figures/Figure03_results_withUG.R",
      "/nature/replication_package/scripts/figures/Figure03_results_withoutUG.R",
      "/nature/replication_package/scripts/figures/Figure03_results_full_country_n500.R",
      "... (44 files omitted) ...",
      "/nature/replication_package/data/figure_and_input_data/two_unit_results_df.csv",
      "/nature/replication_package/data/figure_and_input_data/tc_2km_ug.csv",
      "/nature/replication_package/data/figure_and_input_data/dhs_wi_ug.csv"
    ],
    "execution_steps": [
      "Run the script /nature/replication_package/scripts/figures/Figure03_results_withUG.R",
      "Run the script /nature/replication_package/scripts/figures/Figure03_results_withoutUG.R",
      "Run the script /nature/replication_package/scripts/figures/Figure03_results_full_country_n500.R",
      "Run the script /nature/replication_package/scripts/figures/Figure03_IDW.R",
      "Run the script /nature/replication_package/scripts/figures/Figure03_two_unit_dd_data_creation.R",
      "Run the script /nature/replication_package/scripts/figures/Figure03.R to generate Figure 3"
    ]
  }
}
\end{tcblisting}
\end{tcolorbox}
% \captionof{figure}{Reproduction plan produced by the Setup Agent.}
% \label{fig:plan}
\caption{Reproduction plan produced by the Setup Agent.}
\label{fig:setup}
\end{figure*}

\begin{figure*}[htbp]
\centering
\begin{tcolorbox}[
  colback=white,
  colframe=black,
  title=Execution Summary,
  width=\linewidth,
  boxsep=1mm,
  left=-1mm,right=-1mm,top=-1mm,bottom=-1mm
]
\begin{tcblisting}{
  listing only,
  width=\linewidth,
  boxsep=0mm,
  left=0mm,right=0mm,top=0mm,bottom=0mm,
  sharp corners,
  arc=0mm,
  listing options={
    basicstyle=\ttfamily\footnotesize,
    breaklines=true,
    columns=fullflexible,
    keepspaces=true
  }
}
{
  "code_quality_assessment": "no_errors",
  "reason": "The code logic within the provided R scripts was sound and did not exhibit any code errors. The modifications required were related to adjusting paths for input and output data, which were not logical errors but necessary environmental adjustments. The code executed successfully and produced the expected results, indicating no issues with the inherent code logic or correctness.",
  "Figure 3": {
    "original_files": [
      "/nature/replication_package/scripts/figures/Figure03_results_withUG.R",
      "/nature/replication_package/scripts/figures/Figure03_results_withoutUG.R",
      "/nature/replication_package/scripts/figures/Figure03_results_full_country_n500.R",
      "/nature/replication_package/scripts/figures/Figure03_IDW.R",
      "/nature/replication_package/scripts/figures/Figure03_two_unit_dd_data_creation.R",
      "/nature/replication_package/scripts/figures/Figure03.R"
    ],
    "modified_files": [
      "/nature/replication_package/scripts/figures/Figure03_results_withUG_modified.R",
      "/nature/replication_package/scripts/figures/Figure03_results_withoutUG_modified.R",
      "/nature/replication_package/scripts/figures/Figure03_results_full_country_n500_modified.R",
      "/nature/replication_package/scripts/figures/Figure03_IDW_modified.R",
      "/nature/replication_package/scripts/figures/Figure03_two_unit_dd_data_creation_modified.R",
      "/nature/replication_package/scripts/figures/Figure03_modified.R"
    ],
    "modifications": [
      "Changed data path from relative to absolute",
      "Added functionality to save outputs directly to the figure outputs directory"
    ],
    "output_files": [
      "/nature/Figure03a.pdf",
      "/nature/Figure03b.pdf"
    ]
  }
}
\end{tcblisting}
\end{tcolorbox}
\caption{Execution summary produced by the Execution Agent.}
\label{fig:execution}
\end{figure*}

\begin{figure*}[htbp]
\centering
\begin{tcolorbox}[
  colback=white,
  colframe=black,
  title=Scoring Summary,
  width=\linewidth,
  boxsep=1mm,
  left=-1mm,right=-1mm,top=-1mm,bottom=-1mm
]
\begin{tcblisting}{
  listing only,
  width=\linewidth,
  boxsep=0mm,
  left=0mm,right=0mm,top=0mm,bottom=0mm,
  sharp corners,
  arc=0mm,
  listing options={
    basicstyle=\ttfamily\footnotesize,
    breaklines=true,
    columns=fullflexible,
    keepspaces=true
  }
}
{
  "score": 4,
  "Figure 3": {
    "original_item": "/nature/elements/figure_3_page4.png",
    "reproduced_outputs": [
      "/nature/Figure03a_converted.png",
      "/nature/Figure03b_converted.png"],
    "evaluation_summary": "Figures 3a and 3b were successfully reproduced with matching layout and graphical elements. The numerical patterns and graphical representations closely align with the original figures in the paper."}
}
\end{tcblisting}
\end{tcolorbox}
\caption{Scoring summary produced by the Scoring Agent.}
\label{fig:scoring}
\end{figure*}

\begin{figure*}[htbp]
\centering
\begin{tcolorbox}[
  colback=white,
  colframe=black,
  title=Reproduction Report,
  width=\linewidth,
  boxsep=1mm,
  left=-1mm,right=-1mm,top=-1mm,bottom=-1mm
]
\begin{tcblisting}{
  listing only,
  width=\linewidth,
  boxsep=0mm,
  left=0mm,right=0mm,top=0mm,bottom=0mm,
  sharp corners,
  arc=0mm,
  listing options={
    basicstyle=\ttfamily\footnotesize,
    breaklines=true,
    columns=fullflexible,
    keepspaces=true
  }
}
# Reproducibility Report

## Overall Score
The final reproducibility score is **4**. This indicates that the major findings of the paper, in this case, Figure 3, are fully reproducible.

## Scoring Criteria
- **1**: Major findings in the paper are irreproducible
- **2**: There are minor inconsistencies and/or errors in the provided data and/or code
- **3**: There are rounding errors or equivalent issues in the major findings
- **4**: Major findings of the paper are fully reproducible

## Overall Explanation
The score of **4** was given because Figure 3 was successfully reproduced with no discrepancies between the original and reproduced figures. 
The process involved executing a series of R scripts with the necessary modifications for paths and output customization. The resulting figures 
were visually and numerically consistent with the original representations in the paper. The reproducibility process was straightforward due to 
comprehensive documentation and a well-structured replication package, which facilitated the execution of scripts and verification of outputs. 
Minor modifications were necessary to accommodate file paths and output saving requirements.

## Item-by-Item Analysis

### Figure 3
- **How it was reproduced**: 
  - The following scripts were executed sequentially:
    - `Figure03_results_withUG.R`
    - `Figure03_results_withoutUG.R`
    - `Figure03_results_full_country_n500.R`
    - `Figure03_IDW.R`
    - `Figure03_two_unit_dd_data_creation.R`
    - `Figure03.R`
  - These scripts processed input data files to generate outputs consistent with Figure 3.

- **Modifications made**:
  - Changed data path from relative to absolute.
  - Added functionality to save outputs directly to the figure outputs directory.

- **Output generated**:
  - `Figure03a.pdf`
  - `Figure03b.pdf`

- **Comparison result**:
  - The figures (`Figure03a_converted.png` and `Figure03b_converted.png`) align well with the original Figure 3 from the paper in terms of layout, graphical elements, and numerical data.

- **Reproducibility assessment**:
  - This item was successfully reproduced with all expected elements intact.
\end{tcblisting}
\end{tcolorbox}
\caption{Reproduction report produced by the Report Agent.}
\label{fig:report}
\end{figure*}

% \clearpage

\section{Details of REPRO-Bench-S Construction}
\label{app:repro_bench_s}
\subsection{Instance Correction}
\label{app:repro_bench_correction}
Three authors with at least seven years of programming experience independently review all candidate issues and resolv disagreements through group discussion. We identify 13 problematic instances and apply three types of corrections:
\begin{itemize}
\item[(i)] \textbf{Item removal.} We remove extra reproduction items that are not supported by the reproduction package. This includes items that cannot be linked to any script, intermediate file, or derivable artifact in the repository.

\item[(ii)] \textbf{Scope alignment.} In some instances, the issues documented in the original reproduction report concern results outside the benchmark-specified reproduction items. We revise the ground-truth labels to reflect the reproducibility status of the specified items only.

\item[(iii)] \textbf{Label correction.} We fix factual misannotations in the ground truth. This includes instances marked as reproducible despite missing required data or dependencies, and instances whose reproducibility score conflicts with the evidence in the reproduction report and repository.
\end{itemize}

In total, we identify 13 problematic instances. Among them, 6 are corrected via item removal (Instances 7, 15, 83, 85, 86, 106), 3 via scope alignment (Instances 6, 51, 54), and 4 via label correction (Instances 20, 46, 89, 105). Notably, label changes occur only in scope alignment and label correction, while item removal updates the evaluated items without changing labels.

\subsection{Difficulty Stratification}
\label{app:repro_bench_stratification}
Instances vary substantially in how difficult they are to execute and obtain the required outputs for the reproduction items. We therefore annotate each instance with one of three execution difficulty levels:
\begin{itemize}
\item \textbf{Level-1 (easy).} The repository documentation clearly specifies how to execute the reproduction package to obtain the outputs for each reproduction item. Execution requires editing at most two files, and the edits are limited to minor path fixes. After execution, the artifact(s) for each reproduction item can be identified directly.

\item \textbf{Level-2 (medium).} Instances that do not meet the Level-1 criteria but still provide relatively clear execution guidance. Execution requires editing at most four files. After execution, the artifact(s) corresponding to each reproduction item remain easy to identify.

\item \textbf{Level-3 (hard).} Instances that do not meet the Level-1 or Level-2 criteria. This typically involves unclear execution instructions, more than four files requiring modification, or hard output identification, where execution generates many candidate artifacts, and the system must infer which one(s) correspond to each reproduction item.
\end{itemize}
Three authors labeled all instances independently. We take the majority vote as the final label and resolve ties through discussion to reach consensus.

\end{document}